\documentclass[runningheads,11pt]{mod_llncs}
\usepackage{BTV_09,algorithm,algorithmic}
\usepackage[T1]{fontenc}
\newcommand{\cbw}{{\bf cut\textrm{-}bool}}
\newcommand{\cutbool}{{\bf cut\textrm{-}bool}}
\newcommand{\cutboolw}{{\bf cut\textrm{-}bool\textrm{-}w}}
\begin{document}
\pagestyle{headings}  
\title{Fast FPT algorithms for vertex subset and vertex partitioning problems using neighborhood unions\thanks{Supported by the Norwegian Research Council, project PARALGO.}}
\titlerunning{Fast FPT algorithms for vertex subset and vertex partitioning problems}
\author{B.-M. Bui-Xuan \and J. A. Telle \and M. Vatshelle}
\institute{
Department of Informatics, University of Bergen, Norway.\\
\texttt{[buixuan,telle,vatshelle]@ii.uib.no}
}
\maketitle
\begin{abstract}
We introduce the graph parameter boolean-width, related to the number of different unions of neighborhoods across a cut of a graph.
Boolean-width is similar to rank-width, which is related to the number of $GF[2]$-sums (1+1=0) of neighborhoods instead of the boolean-sums (1+1=1) used for boolean-width.
We give algorithms for a large class of NP-hard vertex subset and vertex partitioning problems that are FPT when parameterized by either boolean-width, rank-width or clique-width, with runtime single exponential in either parameter if given the pertinent optimal decomposition.
To compare boolean-width versus rank-width or clique-width, we first show that for any graph, the square root of its boolean-width is never more than its rank-width.
Next, we exhibit a class of graphs, the Hsu-grids, for which we can solve NP-hard problems in polynomial time, if we use the right parameter.
An $n \times \frac{n}{10}$ Hsu-grid on $\frac{1}{10}n^2$ vertices has boolean-width $\Theta(\log n)$ and rank-width $\Theta(n)$.
Moreover, any optimal rank-decomposition of such a graph will have boolean-width $\Theta(n)$, {\sl i.e.} exponential
in the optimal boolean-width.
A main open problem is to approximate the boolean-width better than
what is given by the algorithm for rank-width of Hlin\v{e}n\'y and Oum \cite{HO08}.
\end{abstract}
%
%
\section{Introduction}
There are several strategies for coping with NP-hard problems, including the use of heuristics, SAT solvers, constraint satisfaction resolution, exact algorithms, approximation algorithms, fixed parameter tractable (FPT) algorithms etc.
In this paper we concentrate on the latter paradigm (see, {\sl e.g.},~\cite{DF99,FG06} for details),
and aim for FPT algorithms with low exponential dependency on the parameter.
We focus on a class of vertex subset and vertex partitioning problems
that include a large number of well-known NP-hard graph problems related to domination, independence,
homomorphism, and so on (see the next section for details).
Algorithms parameterized by either the tree-width~\cite{TP97} of the input graph,
or by its clique-width~\cite{GK03}, have already been given for this class of problems.

Clique-width is one of the graph parameters associated with a powerful framework of decomposition.
The associated FPT algorithms usually have two steps:
\emph{first} compute a decomposition of the input graph,
and \emph{then} solve the in-focus problem along that decomposition.
This relies on the fact that, not only clique-width, but most of the known graph width parameters define decompositions that ease the use of divide-and-conquer techniques, or improvements such as dynamic programming ({\sl e.g.},~\cite{B88,DFT07,GK03,KR03}).
However, as far as clique-width decompositions are concerned, the question of computing an optimal decomposition in FPT time still remains open.

The rank-width parameter consists in a recently introduced framework having the property that an optimal rank-width decomposition can be computed in FPT time~\cite{HO08}.
This is important since an optimal rank-width decomposition can be turned into a $2^{OPT+1}$-approximation of an optimal clique-width decomposition~\cite{OS06}.
This is the first, and still the only, result for approximating optimal clique-width decompositions.
Since the rank-width of a graph is never larger than its clique-width~\cite{OS06}, any problem which is FPT when parameterized by clique-width remains FPT when parameterized by rank-width.
Accordingly, all the work developed for algorithms on graphs of bounded clique-width apply.
For instance, one can combine~\cite{CMR00} and~\cite{HO08} to obtain FPT algorithms, parameterized by either clique-width or rank-width, for any problem expressible in $MSO_1$-logic, given only the graph as input.
However, the main disadvantage when transiting from a rank decomposition to a clique-width decomposition is that the above factor $2^{OPT+1}$ is essentially tight (deducible from~\cite{CR05}). Thus, even if we assume that an optimal rank decomposition is given as input, the runtime to solve NP-hard problems via this transition can unlikely be single exponential in the rank-width\footnote{For a polynomial function $poly$ we call $2^{poly(k)}$ single exponential in $k$, $2^{2^{poly(k)}}$ double exponential in $k$, and so on.}.

This paper is in continuation of the research stream initiated in~\cite{BTV08}, where we gave the first algorithms for NP-hard problems that had runtime single exponential in the rank-width when given an optimal rank-width decomposition.
More recently, an independent work~\cite{GH09} also addresses such questions of single exponential runtimes in the rank-width.
The framework developed in~\cite{GH09} succeeds in giving a positive answer to this question for a number of NP-hard problems including co-Coloring and acyclic-Coloring for a bounded number of colors.
In this paper, when an optimal rank-width decomposition is given, we provide algorithms solving vertex subset and vertex partitioning problems
with runtime single exponential in rank-width, answering an open question raised by~\cite{GH09}.
In terms of FPT time performance, these are improvements of previously known algorithms
by significant orders of magnitude.
This holds even when compared to the case specific algorithms given in~\cite{KR03}.
More comparable with our settings, the algorithms given in~\cite{GK03} address a general class of vertex subset and vertex partitioning problems that is slightly different than ({\sl i.e.} incomparable to) ours.
For those similar to the problems we consider, the algorithms in~\cite{GK03} have runtimes containing a double exponential in the clique-width, {\sl i.e.} a triple exponential in the rank-width.

Basically, the rank-width of a graph $G$ relies on the so-called cut-rank value of a vertex subset $A\subseteq V(G)$.
This is the base $2$ logarithm of the number of subsets of $V(G)\setminus A$ that can be obtained using the symmetric difference of neighborhoods of vertices in $A$ across the cut $\{A,V(G)\setminus A\}$.
It is a matter of fact that, even though they have the best FPT performance so far \emph{when parameterized by rank-width}, the algorithms we give here \emph{do not} manipulate symmetric differences of neighborhoods.
Instead, we manipulate the union of neighborhoods across a cut $\{A,V(G)\setminus A\}$ in a setting similar to rank-width.
For a tighter worst-case complexity analysis, we introduce the notion of boolean-width, obtained by replacing $GF[2]$-sum with boolean-sum in the definition of rank-width.
We give upper and lower bounds linking boolean-width,
on the one hand to rank-width,
on the other hand to clique-width.
We prove that all our bounds are essentially tight for a family of graphs of arbitrary (clique/rank/boolean)-width, except for one, where we leave the tightness as an open question.
In particular, the boolean-width can be as small as the logarithm of rank-width, while the rank-width can never be less than the square root of the boolean-width.
Surprisingly, with respect to clique-width both of them behave similarly, namely we have that
$rw\leq cw\leq 2^{rw+1}-1$~\cite{OS06} and $\beta w\leq cw\leq2^{\beta w+1}$.

Besides this, we exhibit a class of graphs, the so-called Hsu-grids, that will play a role analogous to the role played by cliques in the well-known comparison of clique-width versus tree-width, being a class of graphs for which we can solve NP-hard problems in polynomial time, if we use the right parameter.
An $n \times \frac{n}{10}$ Hsu-grid on $\frac{1}{10} n^2$ vertices has boolean-width $\Theta(\log n)$ and rank-width $\Theta(n)$.
Moreover, any optimal rank-decomposition of such a graph will have boolean-width $\Theta(n)$, {\sl i.e.} exponential
in the optimal boolean-width.
A main open problem is to approximate the boolean-width of a graph better than
what is given by the algorithm for rank-width \cite{HO08}.

Assuming optimal rank-width decomposition $(T, \delta)$ of a graph $G$ given as input, we claim that it is still preferable, for many optimization problems on $G$,
to use boolean-width rather than rank-width in the design and analysis of dynamic programming algorithms based on $(T, \delta)$.
To argue for this note that we prove that if the rank-width of $(T, \delta)$ is $k$ then the boolean-width of $(T, \delta)$ lies between
$\log k$ and $k^2$. Moreover, we believe that the runtime of boolean-width-based algorithms as a function of $k$ would beat or match the best runtimes
of rank-width-based algorithms, for all optimization problems where the number of unions of neighborhoods (boolean sums) across a cut rather than the number of
symmetric differences of neighborhoods (GF[2]-sums) across a cut is the crucial algorithmic bottleneck.

Finally, we remark that the use of boolean-sums instead of GF[2]-sums means a new application for the theory of
boolean matrices, {\sl i.e.} matrices with boolean entries, to the field of algorithms.
Boolean matrices already have applications,
{\sl e.g.} in switching circuits, voting methods, applied logic, communication complexity
and in social science settings like social welfare problems \cite{DKR99,K82}.
A well-studied problem in the field of boolean matrix theory, stemming from a question in
a 1992 paper \cite{K92} concerns the cardinalities of the row spaces of boolean matrices.
This question relates to the
boolean-width parameter which is based on taking the base 2 logarithms
of the cardinalities of such row spaces. In the Conclusion section we indicate how the runtime
of our algorithms could be improved by a positive answer to a question about row spaces of some
(generalized) boolean matrices.

\section{Framework}\label{sec_def}
Let $G$ be a graph with vertex set $V(G)$ and edge set $E(G)$. 
We are interested in the following problems as defined in~\cite{TP97}.
\begin{definition}\label{def_sigma_rho}
\emph{Let $\sigma$ and $\rho$ be finite or co-finite subsets of natural numbers. 
A subset $X$ of vertices of a graph $G$ is a {\sl sigma-rho set}, or simply {\sl $(\sigma,\rho)$-set}, of $G$ if
$$~\hfill\forall v \in V(G): |N(v) \cap X| \in  \left\{ 
   \begin{array}{ll}
    \sigma & \mbox{if $v \in X$}, \\
    \rho & \mbox{if $v \in V(G) \setminus X$}.
   \end{array} \right.\hfill~$$
}
\end{definition}

The {\sl vertex subset problems} consist of finding the size of a minimum or maximum $(\sigma$,$\rho)$-set in $G$. Several NP-hard problems are expressible in this framework, {\sl e.g.},
Max Independent Set, Min Dominating Set, Max Strong Stable Set, Max or Min Perfect Code, 
Min Total Dominating Set,
Max or Min Independent Dominating Set, Min Perfect Dominating Set, Min $k$-Dominating Set, Max Induced $k$-Regular Subgraph,  
Max Induced $k$-Bounded Degree Subgraph
(refer to~\cite{TP97} for further details and a more complete list).
This framework is extendible to problems asking for a partition of $V(G)$ into $q$ classes,
with each class satisfying a certain $(\sigma,\rho)$-property, as follows.
\begin{definition}
\label{def1}
\emph{A {\sl degree constraint} matrix $D_q$ is a $q$ by $q$ matrix with entries being 
finite or co-finite subsets of natural numbers. 
A {\sl $D_q$-partition} in a graph $G$ is a partition $\{V_1,V_2,...,V_q\}$ of 
$V(G)$ such that 
for $1 \leq i,j \leq q$ we have 
$\forall v \in V_i:
|N(v) \cap V_j| \in D_q[i,j]$.
}
\end{definition}

The {\sl vertex partitioning problems} for which we give algorithms in this paper consist of deciding if $G$ has
a $D_q$ partition, the so-called $\exists D_q$ problem.
NP-hard problems fitting into this framework include {\sl e.g.} for any fixed graph $H$
the problems known as $H$-Coloring or $H$-Homomorphism (with $q$-Coloring being $K_q$-Coloring), $H$-Covering, $H$-Partial Covering, and in general the question of deciding if an input
graph has a partition into $q$ $(\sigma, \rho)$-sets, which is in most cases NP-hard already for either $q=2$ or $q=3$~\cite{HT98}
(refer to~\cite{TP97} for further details and a more complete list of problems expressible as $\exists D_q$ problems).
Let us mention that extending the algorithms we give here to handle also the case of finding an extremal value (maximum or minimum) of the cardinality of a vertex partition class over all $D_q$-partitions is quite straightforward.
An efficient technique to solve these problems makes use of dynamic programming on graph decompositions,
done {\sl e.g.} in~\cite{TP97} based on tree decompositions and parameterized by treewidth,
and in~\cite{GK03} based on clique-width expressions and parameterized by clique-width.
With some abuse in terminology, let a subcubic tree be an unrooted tree where all internal nodes have degree three.
\begin{definition}[Decomposition tree]
\emph{A \emph{decomposition tree} of a graph $G$ is a pair $(T,\delta)$ where $T$ is a subcubic tree having $n=|V(G)|$ leaves and $\delta$ is a bijection between the vertices of $G$ and the leaves of $T$.}
\end{definition}

Roughly, trees with their leaves in a bijection with the vertices of $G$ are important for techniques like divide-and-conquer or dynamic programming since they show how to ``divide'' the graph instance into several sub-instances and recurse.
An alternative is to use rooted binary trees instead of subcubic trees, like in clique-width expressions or modular decomposition.
In this paper, we only address unrooted trees, and the question of performing dynamic programming along them.
Clearly, any tree with the right number of leaves and a bijection can be considered as a decomposition tree.
Then, a common technique to select those that are more suited for some task is to use an evaluating function.
A set function $f:2^V\rightarrow\mathbb{R}$ is symmetric if $f(A)=f(V\setminus A)$ for all $A\subseteq V(G)$.
A \emph{cut function} of a graph $G$ is a symmetric function over the vertex subsets of $G$ and will also be seen as a function over all cuts $\{A,V(G)\setminus A\}$ of $G$.
\begin{definition}[Decomposition and width parameters]\label{def_f_width}
\emph{Let $f$ be a cut function of a graph $G$, and $(T,\delta)$ a decomposition tree of $G$.
For every edge $uv$ in $T$, $\{X_u,X_v\}$ denotes the 2-partitions of $V(G)$ induced by the leaf sets of the two subtrees we get by removing $uv$ from $T$.
The \emph{$f$-width of $(T,\delta)$} is the maximum value of $f(X_u)$, taken over every edge $uv$ of $T$.
An \emph{optimal $f$-decomposition} of $G$ is a decomposition tree of $G$ having minimum $f$-width.
The \emph{$f$-width of $G$} is the $f$-width of an optimal $f$-decomposition of $G$.
}
\end{definition}
\begin{remark}
It is clear that an optimal $f$-decomposition of $G$ is not necessarily an optimal $f'$-decomposition of $G$, for $f\neq f'$.
Besides, the symmetry requirement on $f$ is specifically designed for subcubic trees.
This can be relaxed when dealing with rooted binary trees.
\end{remark}

When a decomposition tree of $G$ is given, we provide in Section~\ref{sec_algo} dynamic programming algorithms to solve on $G$ all vertex subset and vertex partitioning problems as defined above.
Roughly, the main idea will be a dynamic classification of the vertex subsets into possible unions -- or boolean sums -- of the neighborhood of some vertices.
A consequence will be that the complexity analysis will be expressed using a function counting the number
of different such neighborhood unions.
Accordingly, we are interested in decomposition trees optimal w.r.t.\ this function.
It is formally defined as follows.
The union of neighborhoods of a vertex subset $X$ is $N(X)=\bigcup_{x\in X}N(x)$.
\begin{definition}[Boolean-width]
\emph{The {\sl boolean-cut function} $\beta_G:2^{V(G)}\rightarrow\mathbb{R}$ of a graph $G$ is defined as\\[6pt]
$~\hfill\beta_G(A)=\log_2 |\{X\subseteq A~:~\exists Y\subseteq B \wedge N(Y)\setminus B=X\}|,\textrm{ where }B=V(G)\setminus A.\hfill~$\\[6pt]
It is known from boolean matrix theory that $\beta_G$ is symmetric~\cite[Theorem 1.2.3]{K82}.
The {\sl boolean-width decompositions and parameters} of $G$ refer to those of Definition~\ref{def_f_width} when $f=\beta_G$.
The $\beta_G$-width of $G$ will be called the \emph{boolean-width} of $G$ and denoted by $\beta w(G)$.
}
\end{definition}

Notice for any graph $G$ on $n$ vertices that $0\leq \beta w(G)\leq n$.
For a vertex subset $A$, the value of $\beta_G(A)$ can also be seen as the logarithm in base $2$ of the number of pairwise different vectors that are spanned -- via boolean sums -- by the rows (resp.\ columns) of the
$A \times V(G) \setminus A$ submatrix of the adjacency matrix of $G$.
Unfortunately, the question of computing an optimal boolean decomposition has not yet been solved.
In order to get an algorithm solving vertex subset and vertex partitioning problems when given only the graph $G$ as input,
we first show that one can approximate an optimal boolean decomposition by a so-called optimal rank decomposition, for which there does exist good algorithms.

\section{Comparing boolean-width to rank-width and clique-width}\label{sec_param}
Rank decompositions and parameters are defined as in Definition~\ref{def_f_width} when the cut function is $\rho_G$, the so-called \emph{cut-rank} function~\cite{O05,OS06}.
\begin{definition}[Rank-width]
\emph{For a vertex subset $A\subseteq V(G)$, $\rho_G(A)$ is defined as the logarithm in base $2$ of the number of pairwise different vectors that are spanned -- via $GF[2]$-sums -- by the rows (resp.\ columns) of the
$A \times V(G) \setminus A$ submatrix of the adjacency matrix of $G$.
The $\rho_G$-width of $G$ will be called the {\sl rank-width} of $G$ and denoted by $rw(G)$.}
\end{definition}

Unlike the boolean-cut function, note that the value of the cut-rank function is always an integer.
For fixed $k$, given a graph $G$ we can in $O(n^3)$ time \cite{HO08} decide if $rw(G) \leq k$ and if so find a decomposition of this width, meaning that the problem of computing rank-width is FPT when parameterized by rank-width.
In this section, we show how one can approximate optimal boolean decompositions with rank decompositions.
More precisely, we will prove for every graph $G$ that
$\log rw(G)\leq \beta w(G)\leq \frac{1}{4}rw(G)^2+\frac{5}{4}rw(G)+\log rw(G),$
with the lower bound being tight to a constant multiplicative factor.
We first investigate the relationship between the cut-rank and the boolean-cut functions.
The following lemma can be derived from a reformulation of~\cite[Proposition 3.6]{BTV08}.
For the sake of completeness, we will give a complete proof using the new terminology.

\begin{lemma}\label{lem_mysterious}\cite[Proposition 3.6]{BTV08}
Let $G$ be a graph and $A\subseteq V(G)$.
Let $nss_G(A)$ be the number of spaces that are $GF[2]$-spanned by the rows (resp.\ columns) of the
$A \times V(G) \setminus A$ submatrix of the adjacency matrix of $G$.
Then, $\log\rho_G(A)\leq\beta_G(A)\leq\log nss_G(A)$.
Moreover, it is well-known from linear algebra that
$nss_G(A)\leq 2^{\frac{1}{4}\rho_G(A)^2+\frac{5}{4}\rho_G(A)}\rho_G(A).$
\end{lemma}
\begin{proof}
This proof is an adaptation of the one given in~\cite[Proposition 3.6]{BTV08} to the terminology used in this paper.
Let $M$ be the $A\times V(G)\setminus A$ submatrix of the adjacency matrix of $G$.
Let $\{a_1,a_2,\dots,a_{\rho_G(A)}\}$ be a set of vertices of $A$ whose corresponding rows in $M$ define a basis of $M$.
Then, it is clear from definition that $N(a_1)$, $N(a_2)$, \dots, $N(a_{\rho_G(A)})$ are pairwise distinct, and from that the first inequality follows, namely $\rho_G(A)\leq2^{\beta_G(A)}$.

Now let $X\subseteq A$.
We define $R_X\subseteq A$ by the following algorithm.\\[6pt]
\indent~~~Initialize $R_X\leftarrow \emptyset$ and $S\leftarrow \emptyset$\\
\indent~~~For every vertex $v\in A$ do:\\
\indent~~~~~~Let $T=N(v)\setminus A$\\
\indent~~~~~~If $T\subseteq N(X)$ and $T\setminus S\neq\emptyset$ then add $v$ to $R_X$ and add all vertices in $T$ to $S$.\\[6pt]
Then, the rows in $M$ which correspond to the vertices of $R_X$ are $GF[2]$-independent.
Hence, every union of neighbourhood of vertices of $A$ can be associated with (at least) one space that is $GF[2]$-spanned by some rows of $M$.
This implies $2^{\beta_G(A)}\leq nss_G(A)$.

For the last inequality, we notice that $$nss_G(A)\leq\sum_{i=1}^{\rho_G(A)}{\rho_G(A) \choose i}_2, \textrm{ where } {n\choose m}_q=\prod_{i=1}^{m}\frac{1-q^{n-i+1}}{1-q^{i}}.$$
This is because ${n\choose m}_q$, which is known under the name of the $q$-binomial coefficient of $n$ and $m$,
is exactly the number of different subspaces of dimension $m$ of a given space of dimension $n$ over a finite field of $q$ elements (roughly, $\frac{1-q^{n-i+1}}{1-q^{i}}$ is the number of choices of an $i^{\textrm{th}}$ vector that is linearly independent from the previously chosen ones).
Now let $a(\rho_G(A))=\sum_{i=1}^{\rho_G(A)}{\rho_G(A) \choose i}_2$.
In order to conclude we can use the $q-$analog of Pascal triangles:
${n\choose m}_q=2^m{n-1\choose m}_q+{n-1\choose m-1}_q,\textrm{ for all }m\leq n,$
with the convention that ${n\choose m}_q=0$ if $m<0$ or $m>n$.
From this we firstly have that the highest number among ${n\choose m}_q$, for all $0\leq m\leq n$, is when $m=\lceil\frac{n}{2}\rceil$.
Therefore,
$a(n)\leq n\times b(n)$ with $b(n)={n\choose \lceil\frac{n}{2}\rceil}_q$.
Finally, still using the $q$-analog of Pascal triangles, one can check that
$b(n)\leq \left(2^{\lceil\frac{n}{2}\rceil}+1\right)\times b(n-1)\leq 2^{\frac{1}{4}n^2+\frac{5}{4}n}$.
\qed
\end{proof}

\begin{figure}[t!]
\centerline{\includegraphics[width=0.675\textwidth]{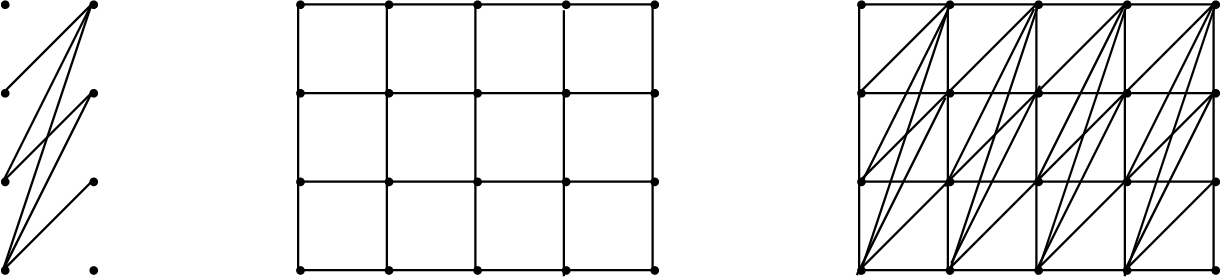}}
\caption{\label{fig_hsu_grid}
The Hsu's graph $H_3$, the $4\times5$ grid, and the Hsu-grid $HG_{4,5}$.}
\end{figure}

We now prove that both bounds given in this lemma are tight.
For the lower bound we recall the graphs used in the definition of Hsu's generalized join~\cite{H87}.
For all $k\geq1$, the graph $H_k$ is defined as the bipartite graph having color classes $A(H_k)=\{a_1,a_2,\dots,a_{k+1}\}$ and $B(H_k)=\{b_1,b_2,\dots,b_{k+1}\}$ such that
$N_{H_k}(a_1)=\emptyset$ and
$N_{H_k}(a_i)=\{b_1,b_2,\dots,b_{i-1}\}$ for all $i\geq2$
(an illustration is given in Figure~\ref{fig_hsu_grid}).
In these graphs, a union of neighborhoods of vertices of $A(H_k)$ is always of the form $\{b_1,b_2,\dots,b_l\}$ with $1\leq l\leq k$, hence,
\begin{lemma}\label{lem_mk_tight}
For $k\geq1$, $\beta_{H_k}(A(H_k))=\log k$ and $\rho_{H_k}(A(H_k))=k$.
\end{lemma}

For the tightness of the upper bound of Lemma~\ref{lem_mysterious} we now recall the graphs used in the characterization of rank-width given in~\cite{BTV08}.
For all $k\geq1$, we denote by $\llbracket1,k\rrbracket=\{1,2,\dots,k\}$.
The graph $R_k$ is defined as a bipartite graph having color classes $A(R_k)=\{a_S,~S\subseteq\llbracket1,k\rrbracket\}$ and $B(R_k)=\{b_S,~S\subseteq\llbracket1,k\rrbracket\}$ such that $a_S$ and $b_T$ are adjacent if and only if $|S\cap T|$ is odd.
\begin{lemma}\label{lem_rk_tight}
For $k\geq1$, $\beta_{R_k}(A(R_k))=\log nss_{R_k}(A(R_k))$ and $\rho_{R_k}(A(R_k))=k$.
\end{lemma}
\begin{proof}
For convenience, we begin with a useful claim.

\medskip

\noindent\textbf{Claim \ref{lem_rk_tight}.1.} \emph{Let $k\geq1$ be an integer.
We keep the notations used in the definition of $R_k$. 
Let $X\subseteq\{a_S,~S\subseteq\llbracket1,k\rrbracket\}$ be such that $a_S,a_T\in X\Rightarrow a_{S\Delta T}\in X$.
Then, we have
$\forall S\subseteq\llbracket1,k\rrbracket,~~N(a_S)\subseteq N(X)\Rightarrow a_S\in X.$}

\medskip

\noindent\textit{Proof.}
Let $M(R_k)$ denote the bipartite adjacency matrix of $R_k$.
It is a straightforward exercice to check that the rows in $M(R_k)$ which correspond to the vertices of $X$ form a $GF[2]$-space of dimension at most $k$ (and also hereafter we will always assume that this exercice is implicit whenever such a vertex subset $X$ is involved).
Besides, in general, a vertex subset $X$ forms a subspace equal to the closure under $\Delta$ (symmetric difference) of the rows corresponding to $X$, i.e. of the neighborhoods of vertices in $X$.
For $R_k$, as opposed for example to Hsu's graph $H_k$, this closure of neighborhoods never goes outside the neighborhoods of $R_k$, because of the following claim, whose proof is straightforward.

~~~~~~~Claim: $N(a_{S}) \Delta N(a_{T}) = N(a_{S \Delta T})$

We denote the dimension of $X$ by $dim(X)$, and prove the lemma by induction on $p=dim(X)$.
If $p=1$ then $X$ contains only one vertex, say $X=\{a_T\}$.
If $S\setminus T\neq\emptyset$, then any element $i\in S\setminus T$ would lead to $b_{\{i\}}$ being a neighbour of $a_S$ but not $a_T$, contradicting that $N(a_S)\subseteq N(X)$.
If $S\subsetneq T$, then any pair $(i,j)$ with $i\in S$ and $j\in T\setminus S$ would lead to $b_{\{i,j\}}$ being a neighbour of $a_S$ but not $a_T$, contradicting that $N(a_S)\subseteq N(X)$.
Hence, $S=T$, which in particular means $a_S\in X$.

We now assume that the lemma is true upto dimension $p-1\geq1$, and consider $X$ such that $dim(X)=p$.
In particular $X$ contains at least two vertices.
Let $a_T$ be a vertex in $X$ such that $a_S\neq a_T$.
If $T\setminus S\neq\emptyset$, we define $W=\{i\}$ with $i\in T\setminus S$, otherwise $T\subsetneq S$ and we define $W=\{i,j\}$ with $i\in S\setminus T$ and $j\in T$.
In any case, we have that $b_W\in N(a_T)\setminus N(a_S)$.
Let $X'=\{a_Z\in X,~b_W\notin N(a_Z)\}=\{a_Z\in X,~|W\cap Z|\textrm{ is even}\}$.
Clearly, $N(a_S)\subseteq N(X')$.
Moreover, we here also have that $a_Z,a_{Z'}\in X'\Rightarrow a_{Z\Delta Z'}\in X'$.
Indeed, since $X'\subseteq X$, we have that $a_{Z\Delta Z'}\in X$.
Besides, it is clear that if both $|W\cap Z|$ and $|W\cap Z'|$ are even, then $|W\cap (Z\Delta Z')|$ is even.
Now, we can conclude by applying the inductive hypothesis on $X'$, which is of dimension lesser than $p-1$.
$\Box$

\medskip

We now prove the lemma by claiming a stronger fact.
Let any subset $Y$ of vertices of $B(R_k)$ be associated with a subset $f(Y)$ of vertices of $A(R_k)$ as $f(Y)=\{a_S,~b_S\notin Y\}$.
Such a set $f(Y)$ can also be seen as a set of rows of $M(R_k)$, the bipartite adjacency matrix of $R_k$.
Then, $f$ is a bijection from the set $UN(R_k)$ of all unions of neighborhoods of some vertices of $A(R_k)$ to the set of all vector spaces spanned by some rows of $M(R_k)$.

Indeed, $f$ is a well-defined function over the subsets of vertices of $B(R_k)$.
It is also clear that $f$ is injective.
We first look at the image of $UN(R_k)$ by $f$.
Let $Y\in UN(R_k)$ and let $X\subseteq\{a_S,~S\subseteq\llbracket1,k\rrbracket\}$ be such that $Y=N(X)$.
Let $a_S,a_T\in f(Y)$.
By definition, neither $b_S$ nor $b_T$ belong to $Y$.
In particular, for every $a_W\in X$, we have that both $|S\cap W|$ and $|T\cap W|$ are even, which also means $|(S\Delta T)\cap W|$ is even.
This implies $b_{S\Delta T}\notin N(X)$.
Hence, $a_{S\Delta T}\in f(Y)$.
In other words, the rows of $M(R_k)$ which correspond to the vertices of $f(Y)$ form a vector space over $GF[2]$.

In order to conclude, we only need to prove that, for every $X\subseteq\{a_S,~S\subseteq\llbracket1,k\rrbracket\}$ such that $a_S,a_T\in X\Rightarrow a_{S\Delta T}\in X$, there exists $Y\in UN(R_k)$ such that $f(Y)=X$.
Let us consider such a set $X$, and define $Y=\{b_S,~a_S\notin X\}$.
It is clear that $f(Y)=X$, and so the only thing left to prove is that $Y\in UN(R_k)$.
More precisely, let $X'=f(N(X))$, we will prove that $Y=N(X')$.
\begin{itemize}
\item Let $b_S\in N(X')$.
Then, there exists $a_T\in X'$ such that $|S\cap T|$ is odd.
By definition of $X'=f(N(X))$, if $a_T\in X'$ then we know that $b_T\notin N(X)$.
Since $S\cap T$ is odd (hence $a_S$ and $b_T$ adjacent in $R_k$), we deduce that $a_S\notin X$.
Then, by definition of $Y$, we deduce that $b_S\in Y$.
Hence, $N(X')\subseteq Y$.
\item Let $b_S\notin N(X')$.
Then, for every $a_T\in X'$, $|S\cap T|$ is even.
Applying the definition of $X'=f(N(X))$, we obtain that, for every $b_T\notin N(X)$, $|S\cap T|$ is even.
In other words, for every $b_T\notin N(X)$, $b_T\notin N(a_S)$.
Therefore, $N(a_S)\subseteq N(X)$.
The above Claim \ref{lem_rk_tight}.1 then applies and yields $a_S\in X$.
This, by definition of $Y$, means $b_S\notin Y$.
Hence, $Y\subseteq N(X')$.
\end{itemize}
\qed
\end{proof}

Since Lemma~\ref{lem_mysterious} holds for all edges of all decomposition trees, it is clear that
for all graphs $G$ we have
$\log rw(G)\leq \beta w(G)\leq \frac{1}{4}rw(G)^2+\frac{5}{4}rw(G)+\log rw(G).$
We now address the tightness of this lower bound.
We say that a cut $\{A,V(G)\setminus A\}$ is \emph{balanced} if $\frac{1}{3}|V(G)|\leq|A|\leq\frac{2}{3}|V(G)|$.
In any decomposition tree of $G$, there always exists an edge of the tree which induces a balanced cut in the graph.
We lift the tightness result on graph cuts given by Lemma~\ref{lem_mk_tight} to the level of graph parameters in a standard way, by using the structure of a grid as follows.
The main idea is that any balanced cut of a grid will contain either a large part of some column of the grid, or it contains a large enough matching.
We then add edges to the columns of the grid and fill each of them into a Hsu graph.
The formal definition is given below while an illustration is given in Figure~\ref{fig_hsu_grid}.
Note that graphs with a similar definition have also been studied in relation with clique-width in a different context~\cite{BL03}.
\begin{definition}[Hsu-grid $HG_{p,q}$]
\emph{Let $p\geq2$ and $q\geq2$.
The Hsu-grid $HG_{p,q}$ is defined by
$V(HG_{p,q})=\{v_{i,j}~|~1\leq i\leq p~\wedge~1\leq j\leq q\}$ with
$E(HG_{p,q})$ being exactly the union of the edges $\{(v_{i,j},v_{i+1,j})~|~1\leq i<p~\wedge~1\leq j\leq q\}$ and of the edges $\{(v_{i,j},v_{i',j+1})~|~1\leq i\leq i'\leq p~\wedge~1\leq j<q\}$.
We say that vertex $v_{i,j}$ is at the $i^{th}$ row and the $j^{th}$ column.
}
\end{definition}
\begin{lemma}\label{main_lem_hsu_grid}
For large enough integers $p$ and $q$, we have that
$\beta w(HG_{p,q})\leq\min(2\log p,q)$ and $rw(HG_{p,q})\geq\min(\lfloor\frac{p}{4}\rfloor,\lfloor\frac{q}{6}\rfloor)$.
Moreover, if $q<\lfloor\frac{p}{8}\rfloor$ then any optimal rank decomposition of $HG_{p,q}$ has boolean-width at least $\lfloor\frac{q}{6}\rfloor$.
\end{lemma}
\begin{proof}
Let $m/n$ denote $\lfloor\frac{m}{n}\rfloor$.
We begin with a useful claim.

\medskip

\noindent\textbf{Claim \ref{main_lem_hsu_grid}.1.}
\emph{Let $p\geq2$ and $q\geq2$.
Let $\{A,B\}$ be a balanced cut of the Hsu-grid $HG_{p,q}$, and let $H$ be the bipartite graph containing all edges of $HG_{p,q}$ crossing $\{A,B\}$.
Then, either the cut-rank of $A$ is at least $p/4$, or $H$ contains a $q/6$-matching as induced subgraph.}

\medskip

\noindent\textit{Proof.}
We distinguish two self-exclusive cases.
\begin{itemize}
\item Case 1: for every row $1\leq i\leq p$ there exists an edge $(v_{i,j},v_{i,j+1})$ crossing $\{A,B\}$
\item Case 2: there is a row $1\leq i\leq p$ containing only vertices of one side of the cut, w.l.o.g.\ $v_{i,j}\in A$ for all $1\leq j\leq q$
\end{itemize}

In case 1, we can suppose w.l.o.g.\ that there are at least $p/2$ row indices $i$'s for which there exists $j$ such that $v_{i,j}\in A$ and $v_{i,j+1}\in B$.
Therefore, there are at least $p/4$ row indices $i$'s for which there exists $j$ such that $v_{i,j}\in A$ and $v_{i,j+1}\in B$ and that no two rows among those are consecutive (take every other row).
Now we can check that the rank of the bipartite adjacency matrix of the subgraph of $H$ that is induced by the $p/4$ above mentioned pairs $v_{i,j}$ and $v_{i,j+1}$ is at least $p/4$.
Hence, the cut-rank of $A$ is at least $p/4$.

In case 2, from the balanced property of the cut $\{A,B\}$ we have that there are at least $q/3$ columns each containing at least one vertex of $B$.
Then, for each such column $j$ we can find an edge $(v_{i,j},v_{i+1,j})$ crossing $\{A,B\}$.
Choosing one such edge every two columns will lead to a $q/6$ matching that is an induced subgraph of $H$.
$\Box$

\medskip

To prove the lemma, we will focus on two types of decomposition trees.
A star is a tree having one and only one internal node, the so-called center of the star.
The $2$-leaf comb is the $2$-leaf star.
The $n$-leaf comb is obtained by appending one leaf of the $2$-leaf comb to the center of an $(n-1)$-leaf comb; the center of the $n$-leaf comb is the center of the former comb.
In a tree, \emph{stretching} an internal node $n$ is the action of adding a new node $n'$ adjacent to $n$, and removing at least one neighbour of old node $n$ to be neighbour of new node $n'$.

The vertical super-decomposition tree of the Hsu-grid $HG_{p,q}$ is obtained by appending every leaf of a $q$-leaf comb to the center of a new $p$-leaf star,
unrooting the obtained graph,
and subsequently mapping the $j^{th}$ leaf set of the $q$ different $p$-leaf stars to the set of vertices of $HG_{p,q}$ standing on the $j^{th}$ column in $V(HG_{p,q})$.
A {\sl vertical decomposition tree} is obtained by consecutively stretching all internal nodes of degree more than $3$ in the vertical super-decomposition tree.
The notion of a {\sl horizontal (super-)decomposition tree} is defined similarly when swapping the roles of $p$ and $q$, and that of columns and rows.
Notice from $p\geq3$ and $q\geq3$ that both vertical and horizontal decomposition tree are indeed decomposition tree (in particular their underlying trees are subcubic).

We now come to the actual proof of the lemma.
It is straightforward to check that the boolean-width of any vertical decomposition tree of $HG_{p,q}$ is at most $2\log p$
and the boolean-width of any horizontal decomposition tree of $HG_{p,q}$ is at most $q$.
Therefore, $\beta w(HG_{p,q})\leq\min(2\log p,q)$.
Besides, it follows directly from the above Claim \ref{main_lem_hsu_grid}.1 that $rw(HG_{p,q})\geq\min(p/4,q/6)$.

To prove the third and last claim, we first notice that any horizontal decomposition tree of $HG_{p,q}$ has rankwidth $2q$, and therefore the rank-width of $HG_{p,q}$ is at most $2q<p/4$.
We now consider an optimal rank decomposition of $HG_{p,q}$.
Let $uv$ be an edge of the decomposition which induces a balanced cut $\{A,B\}$ in $HG_{p,q}$.
Let $H$ be the bipartite graph containing all edges of $HG_{p,q}$ crossing $\{A,B\}$.
From the above Claim \ref{main_lem_hsu_grid}.1 and the fact that the rank-width of $HG_{p,q}$ is at most $2q<p/4$,
$H$ has a $q/6$-matching as induced subgraph.
Therefore, the boolean-cut of $A$ is at least $q/6$, and the boolean-width of this optimal rank decomposition is at least $q/6$.
\qed
\end{proof}

Notice that not only the lemma addresses the tightness of the lower bound on boolean-width as a function of rank-width,
but also the additional stronger property that for a special class of Hsu-grids {\sl any} optimal rank decomposition has boolean-width
exponential in the optimal boolean-width. 
This points to the importance of finding optimal boolean decompositions to achieve fast runtimes on these graphs.
\begin{theorem}\label{theo_lw_rw}
For any graph $G$ it holds that
$\log rw(G)\leq \beta w(G)\leq \frac{1}{4}rw(G)^2+\frac{5}{4}rw(G)+\log rw(G)$.
Moreover, for large enough integer $k$, there are graphs $L_k$ and $U_k$ of rank-width at least $k$ such that
$\beta w(L_k)\leq 2\log rw(L_k)+4$ and $\beta w(U_k)\geq \lfloor\frac{1}{6}rw(U_k)\rfloor-1$.
\end{theorem}
\begin{proof}
The bounds for arbitrary graphs follow directly from Lemma~\ref{lem_mysterious}.
We define $L_k$ as a Hsu-grid $HG_{p,q}$ such that $k\leq {p}/{4}\leq {q}/{6}$ and $2\log p\leq q$.
Then, from Lemma~\ref{main_lem_hsu_grid}, we have that $rw(L_k)\geq p/4\geq k$ and $\beta w(L_k)\leq2\log p$, which allows to conclude about $L_k$.
Finally, we define $U_k$ to be the grid of dimension $k\times k$.
It is a standard exercice to check that the rank-width of $U_k$ is at most $k+1$.
The same idea as in the proof of Claim~\ref{main_lem_hsu_grid}.1 (inside the proof of Lemma~\ref{main_lem_hsu_grid}) can be used to prove that $\beta w(U_k)\geq k/6$.
\qed
\end{proof}

\begin{remark}
The inequalities about $L_k$ is a direct application of Lemma~\ref{main_lem_hsu_grid} for well-chosen values of $p$ and $q$.
We leave open the question whether the bound $\beta w(G)\leq\frac{1}{4}rw(G)^2+\frac{5}{4}rw(G)+\log rw(G)$ is tight to a multiplicative factor.
\end{remark}
\begin{remark}\label{rem_crazy_approx}
Let $(T,\delta)$ and $(T',\delta')$ be such that
$rw(G)=rw(T,\delta)$ and $OPT=\beta w(G)=\beta w(T',\delta')$.
We then have from Lemma~\ref{lem_mysterious} that
$\beta w(T,\delta)\leq rw(T,\delta)^2\leq rw(T',\delta')^2\leq(2^{OPT})^2$.
Hence, any optimal rank decomposition of $G$ is also a $2^{2*OPT}$-approximation of an optimal boolean decomposition of $G$.
\end{remark}

We have seen in this section that rank-width can be used to approximate boolean-width, though the gap between them can be quite large, in particular boolean-width can be much lower than rank-width.
One of the most important applications of rank-width is to approximate the so-called clique-width $cw(G)$ of a graph\footnote{A proper introduction to clique-width is beyond the purposes of this paper.
The formalism needed for the proof of Theorem~\ref{theo_lw_cw} will be recalled inside its proof.}~\cite{OS06}.
The bounds are
$\log(cw(G)+1)-1\leq rw(G)\leq cw(G)$~\cite{OS06}.
Although we have seen that the difference between rank-width and boolean-width can be quite large, we remark that, w.r.t.\ clique-width, boolean-width behaves similarly as rank-width, namely
\begin{theorem}\label{theo_lw_cw}
For any graph $G$ it holds that
$\log cw(G)-1\leq \beta w(G)\leq cw(G)$.
Moreover, for every integer $k$ higher than some constant, there are graphs $L_k$ and $U_k$ of clique-width at least $k$ such that
$\beta w(L_k)\leq 2\log cw(L_k)+4$ and $\beta w(U_k)\geq\lfloor\frac{1}{6}cw(U_k)\rfloor-1$.
\end{theorem}
\begin{proof}
We first recall the definition of clique-width.
Using abusive notations, we confound graphs and vertex-labelled graphs (for more concision refer to~\cite{CMR00}).
The \emph{clique-width} of a graph $G$ is the minimum number of labels needed to construct $G$ using the following four operations:
\begin{itemize}
\item Creation of a new vertex $v$ with label $i$ (denoted by $i(v)$).
\item Disjoint union of two labeled graphs $G$ and $H$ (denoted by $G \oplus H$).
\item Joining by an edge each vertex with label $i$ to each vertex with label $j$ ($i\neq j$, denoted by $\eta_{i,j}$).
\item Renaming label $i$ to $j$ (denoted by $\rho_{i\rightarrow j}$).
\end{itemize}

The construction tree of $G$ from these operations is called an \emph{$k$-expression} of $G$, where $k$ is the number of distinct labels needed in the construction.

We continue with a useful definition.
Let $A\subseteq V(G)$ be a vertex subset of $G$.
An \emph{external module partition} of $A$ is a partition $P$ of $A$ such that, for every
$z \in V(G) \setminus A$ and pair of vertices $x,y$ belonging to the same class in $P$, we have $x$ adjacent to $z$ if and only if $y$ adjacent to $z$.
For any $A$, an maximum (coarse-wise) external module partition of $A$ always exists~\cite[Lemma 3.2]{BTV08}.

To prove the lower bound of the theorem, we consider an optimal boolean-width decomposition of $G$, subdivide any edge and obtain a rooted binary tree.
At any internal node $w$ of the tree with children $a$ and $b$,
we denote the vertex subsets induced by the leaves of the subtree rooted at $w$, $a$ and $b$ 
by $V_w$, $V_a$ and $V_b$, respectively.
We also denote the number of parts in the maximum external module partition of $V_w$, $V_a$ and $V_b$ by $k_w$, $k_a$ and $k_b$, respectively.
Clearly, if $G[V_a]$ and $G[V_b]$ have $k_a$-expression and $k_b$-expression, respectively, then $G[V_w]$ has a $(k_a+k_b)$-expression.
Besides, it follows directly from definition of the boolean-cut function that
$k_a\leq2^{\beta_G(V_a)}$ and
$k_b\leq2^{\beta_G(V_b)}$.
From those observations we can deduce by a standard induction that there exists a $2^{\beta w(G)+1}$-expression of $G$.

To prove the upper bound, we consider an $k$-expression of $G$ of optimal clique-width, and contract all internal nodes of degree $2$ (corresponding to the $\rho_{i\rightarrow j}$ operations) in order to obtain a rooted binary tree $T$.
Now, at any edge $uv$ of the tree, with $u$ being the parent of $v$, the number of parts in the maximum external module partition of $V_v$ is no more than $k$, where $V_v$ is the vertex subset induced by the leaves of the subtree rooted at $v$.
Hence, ${\beta_G(V_v)}\leq k$.
Whence, unrooting $T$ results in a decomposition tree having a boolean-width at most $k$.

We now use the same graphs $L_k$ and $U_k$ as in the proof of Theorem~\ref{theo_lw_rw} and the well-known fact that $rw(G)\leq cw(G)$ for any graph $G$~\cite{OS06} in order to conclude that $\beta w(L_k)\leq 2\log cw(L_k)+4$.
It is a standard exercice to check that the cliquewidth of $U_k$ is at most $k+2$.
Combining this with $\beta w(U_k)\geq k/6$ allows to conclude.
\qed
\end{proof}

We now show how to use optimal boolean decompositions -- or their approximations by rank decompositions
which are actually computable \cite{HO08} -- to solve vertex subset and vertex partitioning problems efficiently.
We will also discuss the computational advantages we would get if we could use optimal boolean decompositions instead of optimal rank decompositions.

\section{Algorithms}\label{sec_algo}
Let a graph $G$ and a decomposition tree $(T,\delta)$ of $G$ be given as input.
We consider in the upcoming section the problem of computing the size of a minimum or maximum ($\sigma$,$\rho$)-set in $G$, 
the so-called vertex subset problems (see Definition~\ref{def_sigma_rho}).
The subsequent section is devoted to vertex partitioning problems (see Definition~\ref{def1}).
All solutions we give will follow the dynamic programming algorithmic framework (refer to, {\sl e.g.},~\cite{CLRS2001} for details).

To this aim, we subdivide an arbitrary edge of $T$ to get a new root node $r$, 
and denote by $T_r$ the resulting rooted tree.
The algorithms will follow a bottom-up traversal of $T_r$.
With each node $w$ of $T_r$ we associate a table data structure $\mathit{Tab}_w$, 
that will store optimal solutions to subproblems related to $V_w$,
the set of vertices of $G$ mapped to the leaves of the subtree of $T_r$ rooted at $w$.
From the perspective of an efficient runtime, each index of the table will be associated with a certain class of equivalent subproblems 
that we need to define depending on the problem on which we are focusing.

\subsection{Dynamic programming for vertex subset problems}\label{sec_sigma_rho}
Let $d(\mathbb{N}) = 0$.
For every finite or co-finite set $\mu \subseteq \mathbb{N}$, let $d(\mu) = 1+min ( max_{x \in \mathbb{N}} x : x \in \mu, max_{x \in \mathbb{N}} x: x \notin \mu )$.
We denote by $d(\sigma,\rho)$, or simply by $d$ when it appears clearly in the context that $\sigma$ and $\rho$ are involved, the value $d=d(\sigma,\rho)=max(d(\sigma),d(\rho))$.
Note that when checking if a subset $A$ of vertices is a $(\sigma, \rho)$-set, as in Definition~\ref{def_sigma_rho}, it suffices to
count the number of neighbors up to $d$ that a vertex has in $A$. This is the key to 
getting fast algorithms and is captured by the following equivalence classes.
\begin{definition}[$d$-neighbor equivalence]\label{dNeighbour}
\emph{Let $G$ be a graph and $A\subseteq V(G)$ a vertex subset of $G$. Denote by $\overline{A}$
the vertex set $V(G) \setminus A$.
Two vertex subsets $X \subseteq A$ and $X' \subseteq A$ are {\sl $d$-neighbor equivalent} w.r.t.\ $A$, denoted by $X \equiv^d_A X'$, if\\
$~\hfill\forall v \in \overline{A} : \left(|N(v) \cap X|\frac{}{}=\frac{}{}|N(v) \cap X'|\right) \vee \left(|N(v) \cap X| \geq d\frac{}{} \wedge\frac{}{} |N(v) \cap X'| \geq d \right).\hfill~$.\\
Let $nec(\equiv^d_A)$ be the number of equivalence classes of $\equiv^d_A$.}
\end{definition}
\begin{definition} 
\label{srDomination}
\emph{Let $G$ be a graph, $A\subseteq V(G)$, and $\mu\subseteq\mathbb{N}$.
For $X \subseteq V(G)$, set $X$ $\mu$-dominates $A$ if
$\forall v \in A : |N(v) \cap X| \in \mu$.
For $X \subseteq A$, $Y \subseteq \overline{A}$,
the pair $(X,Y)$ {\sl $\sigma,\rho$-dominates} $A$ if 
$(X \cup Y)\ \sigma$-dominates $X$ and $(X \cup Y)\ \rho$-dominates $A \setminus X$.}
\end{definition}

For simplicity, consider first the effect of simply using any $X \subseteq A$ and $Y \subseteq \overline{A}$ and defining
$Tab_w[X][Y] \stackrel{\mathrm{def}}{=}opt_{S \subseteq V_w} \{ |S| : S \equiv^d_{V_w} X \mbox{ and } 
(S,Y) \mbox{ $\sigma,\rho$-dominates $V_w$}\}$, where $opt$ stands for function $min$ or $max$, accordingly.
Then, at the root $r$ of $T_r$ the value of $Tab_r[X][\emptyset]$ (for all $X\subseteq V(G)$) would be exactly 
equal to the size of a maximum, resp. minimum, $(\sigma,\rho)$-set of $G$ (c.f.\ $\equiv^d_{V_r}$ has only one equivalence class).
Accordingly, a bottom-up computation of $\mathit{Tab}_r$ would be sufficient to solve the vertex subset problem.
However, the sizes of the tables would then depend on $|V(G)|$.
We first show that using $d$-neighbor equivalence classes we lower the size of the tables to a function not depending on $|V(G)|$. 
Clearly, if $X\equiv^d_{V_w}X'$, then the values of $Tab_w[X][Y]$ and $Tab_w[X'][Y]$ are the same (for all $Y$).
Hence, for each equivalence class of $\equiv^d_{V_w}$, we only need to keep one representative entry in $Tab_w$ for all the 
members of that class, thus reducing the number of entries in $Tab_w$ to the number of equivalence classes of $\equiv^d_{V_w}$ 
times the number of equivalence classes of $\equiv^d_{\overline{V_w}}$.
Recall that  $\overline{V_w}$ denotes $V(G)\setminus V_w$.

\begin{lemma}\label{repAlg}
Let $G$ be a graph and $A\subseteq V(G)$.
Then, for every $X\subseteq A$, there is $R \subseteq A$ such that $R \equiv^d_{A} X$ and $|R| \leq d \cdot \cbw(A)$.
Moreover, $nec(\equiv^d_{A}) \leq 2^{d \cdot \cbw(A)^2}$.
\end{lemma}
\begin{proof}
We start with the first part namely bounding the size of the minimal members.
Let $R = X $, go through all vertices $x \in X$ and delete $x$ from $R$ if $R \setminus \{x\} \equiv_A^d X$.
Notice that $R \subseteq X$ and $R \equiv^d_{A} X$ and hence fulfill two of the requirements.

We then know that $\forall x \in R\ \exists y \in N(x) \setminus A : |N(y) \cap X| \leq d$,
this since otherwise $R \setminus \{x\} \equiv_A^d X$.
We build a set $S$ as follows:\\
Let $S = \emptyset$, $R'=R$.
While we can find a pair $x \in R',y \in \overline{A}$ such that $|N(y) \cap R'| \leq d$ and $x \in N(y)$.
Remove $N(y)$ from $R'$, add $x$ to $S$.

Note that all the $x$'s and $y$'s chosen are different since we remove the neighborhood
of $y$ from $R'$ and that we can always find such a set unless $R' = \emptyset$.
Therefore we know $|S| \ge |R|/d$ since we remove at most $d$ nodes in each step.
Any of the $2^ {|S|}$ combinations of elements from $S$ will form a unique neighborhood.
Therefore, we get from definition $\cutbool(A) = \cutbool(\overline{A}) \ge |S|$.
Since $|S| \ge |R|/d$, $R$ fulfills the last requirement.

To bound the number of equivalence classes $nec(\equiv^d_{A})$ we know from the previous arguments that
we only need to find the equivalence classes among the subsets of $A$ of size at most $d \cdot \cutbool(A)$.
Let $H$ be obtained from the bipartite subgraph of $G$ with color classes $A,\overline{A}$ after doing twin contraction of all twins and adding an isolated vertex to each color class unless there already is one.
We know that every node of $V(H) \cap A$ has a unique neighborhood,
 hence $|V(H)\cap A| \leq 2^{\cutbool(A)}$.
Any subset of $A$ is a multiset of $|V(H)\cap A|$, and a trivial bound on
 number of multisets of $|V(H)\cap A|$ with size $d \cdot \cutbool(A)$ gives us:
  $nec(\equiv^d_{A}) \leq 2^{d \cdot \cutbool(A)^2}$.
\qed
\end{proof}

We now define the canonical representative $can^d_{V_w}(X)$ of every subset $X\subseteq V_w$,
and the canonical representative $can^d_{\overline{V_w}}(Y)$ of every subset $Y\subseteq \overline{V_w}$.
For simplicity we define this using $V_w$ instead of a generic subset $A$, but note that everything we say about $X \subseteq V_w$,
$can^d_{V_w}(X)$ and $\equiv^d_{V_w}$ will hold also for $can^d_{\overline{V_w}}(Y)$, $Y\subseteq \overline{V_w}$ and
$\equiv^d_{\overline{V_w}}$.
Canonical representatives are to be used for
indexing the table $Tab_w$ at node $w$ of the tree $T_r$.
Three properties will be required.
Firstly, if $X\equiv^d_{V_w}X'$, then we must obviously have $can^d_{V_w}(X)=can^d_{V_w}(X')$.
Secondly, given $(X,Y)$, we should have a fast routine that outputs a pointer to the entry $Tab_w[can^d_{V_w}(X)][can^d_{\overline{V_w}}(Y)]$.
Thirdly, we should have a list whose elements can be used as indices of the table, {\sl i.e.} a list containing all canonical representatives w.r.t.\ $\equiv^d_{V_w}$.
Let an arbitrary ordering of the vertices of $V(G)$ be given.
The following definition trivially fulfills the first requirement.

\begin{definition}
\emph{We assume that a total ordering of the vertices of $V(G)$ is given.
For every $X \subseteq V_w$, the canonical representative $can^d_{V_w}(X)$ is defined as the lexicographically smallest set $R \subseteq V_w$ such that: 
$|R|$ is minimized and $R \equiv^d_{V_w} X$.
}
\end{definition}

\begin{definition}
\emph{Let $opt$ stand for either function $max$ or function $min$, depending on whether we are looking for a maximum
or minimum $(\sigma,\rho)$-set, respectively.
For every node $w$ of $T_r$, for $X\subseteq V_w$ and $Y\subseteq \overline{V_w}$,
let $R_X=can^d_{V_w}(X)$ and $R_Y=can^d_{\overline{V_w}}(Y)$.
We define the contents of $\mathit{Tab}_w[R_X][R_Y]$ as:
$$ Tab_w[R_X][R_Y] \stackrel{\mathrm{def}}{=}
\left\{\begin{array}{l}opt_{S \subseteq V_w} \{ |S| : S \equiv^d_{V_w} X \mbox{ and } (S,Y) \mbox{ $\sigma,\rho$-dominates $V_w$} \},\\
-\infty\textrm{ if no such set $S$ exists and $opt=max$,}\\
+\infty\textrm{ if no such set $S$ exists and $opt=min$.}
\end{array}\right.$$
}
\end{definition}

\begin{definition}[d-neighborhood]
\label{def_dneigborhood}
For a set $S \subseteq A$, the $d$-neighborhood of $S$, denoted $N^d_A(S)$, is a multiset of nodes from $\overline{A}$, 
such that, $\forall v \in \overline A$ the number of occurrences of $v$ in $N^d_A(S)$ is equal to $\min \{|N(v) \cap S|,d\}$.
Since we have assumed a fixed ordering of the vertices we will store such a multiset as an $|\overline A|$-vector of integers from the interval $[0,d]$.
\end{definition}

\begin{algorithm}[List of representatives and their d-neighborhood]
\caption{List of representatives and their d-neighborhood}
\label{algo_replist}
\begin{algorithmic}
 \STATE INPUT: bipartite graph $G(A,\overline A)$ and integer $d$
 \STATE Initialize $LR_A$, $LNR_A$, $NextLevel$ to be empty
 \STATE Initialize $LastLevel = \{\emptyset\}$
 \WHILE{LastLevel != $\emptyset$}
   \FOR{$R$ in LastLevel}
     \FOR{every vertex $v$ of $A$}
       \STATE $R' = R \cup \{v\}$\\
       \STATE compute $N' = N^d_A(R')$
       \IF{$R' \not\equiv^d_A R$ and $N'$ is not contained in $LNR_A$}
         \STATE add $R'$ to both $LR_A$ and $NextLevel$\\
         \STATE add $N'$ to $LNR_A$ at the proper position\\
         \STATE add pointers between $R'$ and $N'$\\
       \ENDIF
     \ENDFOR
   \ENDFOR
   \STATE set $LastLevel = NextLevel$, and $NextLevel = \emptyset$\\
 \ENDWHILE
 \STATE OUTPUT: $LR_A$ and $LNR_A$
\end{algorithmic}
\end{algorithm}

\begin{lemma}\label{lem_list_cano_rep}
For any node $w$ of $T_r$ with $k = \cbw(V_w)$, we can compute a list containing all representatives w.r.t.\ $\equiv^d_{V_w}$ in time $O(nec(\equiv^d_A) \cdot \log (nec(\equiv^d_A)) \cdot n^2$.
For any subset $X \subseteq V_w$, a pointer to its unique representative in the list of representatives can be found in time $O( \log (nec(\equiv^d_A)) \cdot |X| \cdot n)$.
\end{lemma}
\begin{proof}
 Algorithm \ref{algo_replist} computes such a list.
 Before adding a representative $R$ to the list $LR_{V_w}$ we check if the list $LNR_{V_w}$ contains the d-neighborhood $N^d_{V_w}(R)$. 
 Therfore all elements of the list $LR_{V_w}$ have different d-neighbourhoods.
 All the representatives added to the list $LR_{V_w}$ are also expanded by any of the vertices of ${V_w}$.
 Assume for contradiction that $X$ is a minimal representative such that $N^d_{V_w}(X)$ is not in the list $LNR_{V_w}$.
 Then $\forall u \in X$ we have: 
 $\forall Y \in LNR_{V_w} : X \setminus u \not\equiv Y$ since then $N^d_{V_w}(X)$ would have been added to $LNR_{V_w}$.
 Meaning that $N^d_{V_w}(X \setminus u)$ is not in $LNR_{V_w}$ contradicting that $X$ is minimal.

 The total number of representatives to be added to $LR_{V_w}$ and $d$-neighborhoods added to $LNR_{V_w}$ is $nec(\equiv^d_A)$.
 The total number of possible representatives $R'$ to be considered is $nec(\equiv^d_A) \cdot n$.
 Computing the union $R \cup \{v\}$ and the $d$-neighbouthood $N^d_{V_w}(R')$ can be done in $O(n)$ time 
 by copying the $d$-neigborhood vector of $R$ and updating the entries for vertices in $N(v) \cap \overline{V_w}$.
 If we realize the list $LNR_{V_w}$ as a balanced binary searchtree checking for containment can be done in $O(\log (nec(\equiv^d_A)) \cdot n)$.
 Inserting into the list $LR_{V_w}$ can be done in constant time.
 So in total the construction of $LR_{V_w}$ and $LNR_{V_w}$ takes $O(nec(\equiv^d_A) \cdot \log (nec(\equiv^d_A)) \cdot n^2$.
 
 Given a subset $X \subseteq V_w$ we can generate the $d$-neighborhood $N^d_{V_w}$ in $O(|X| \cdot n)$ time.
 Then we can binarysearch in the list $LNR_{V_w}$ to find a pointer to the representative in time $O( \log (nec(\equiv^d_A)) \cdot |X| \cdot n)$. 
\qed
\end{proof}

Note that at the root $r$ of $T_r$ the value of $Tab_r[X][\emptyset]$ (for all $X\subseteq V(G)$) would be exactly equal to the size of a maximum, resp. minimum, $(\sigma,\rho)$-set of $G$
({\sl cf.} $\equiv^d_{V_r}$ has only one equivalence class).
For initialization, the value of every entry of $Tab_w$ will be set to $+\infty$ or $-\infty$ depending on whether we are solving a minimization or maximization problem, respectively.
For a leaf $l$ of $T_r$, we perform a brute-force update:
let $A=\{l\}$ and $B=\overline{A}$,
for every canonical representative $R$ w.r.t.\ $\equiv^d_B$, we set:
\begin{itemize}
\item If $|N(l)\cap R|\in\sigma$ then $Tab_l[A][R]=1$.
\item If $|N(l)\cap R|\in\rho$ then $Tab_l[\emptyset][R]=0$.
\end{itemize}

For a node $w$ of $T_r$ with children $a$ and $b$, the algorithm proceeds as follows.
For every canonical representative $R_{\overline{w}}$ w.r.t.\ $\equiv^d_{\overline{V_w}}$,
for every canonical representative $R_a$ w.r.t.\ $\equiv^d_{V_a}$,
and for every canonical representative $R_b$ w.r.t.\ $\equiv^d_{V_b}$, do:
\begin{itemize}
\item Compute $R_w = can^d_{V_w}(R_a \cup R_b)$,
$R_{\overline{a}} = can^d_{\overline{V_a}}(R_b \cup R_{\overline{w}})$ and 
$R_{\overline{b}} = can^d_{\overline{V_b}}(R_a \cup R_{\overline{w}})$
\item Update $\mathit{Tab}_w[R_w][R_{\overline{w}}] = opt(\mathit{Tab}_w[R_w][R_{\overline{w}}],\mathit{Tab}_a[R_a][R_{\overline{a}}] + \mathit{Tab}_b[R_b][R_{\overline{b}}]).$
\end{itemize}

\begin{lemma}\label{lem_correctness_sigma_rho}
The table at node $w$ is updated correctly, {\sl i.e.} for any canonical representatives $R_w$ and $R_{\overline{w}}$ w.r.t.\ $\equiv^d_{V_w}$ and $\equiv^d_{\overline{V_w}}$,
if $Tab_w[R_w][R_{\overline w}]$ is not $\pm\infty$ then\\[6pt]
\centerline{$\mathit{Tab}_w[R_w][R_{\overline{w}}] = opt_{S \subseteq V_w}\{ |S| : S \equiv_{V_w}^dR_w  \wedge  (S,R_{\overline w})$ $\sigma,\rho$-dominates $V_w\}.$}\\[6pt]
If the value of the table is $\pm\infty$ then there is no such above set $S$.
\end{lemma}
\begin{proof}
Let $a,b$ be the children of $w$ in $T_r$, assume $\mathit{Tab}_a$ and $\mathit{Tab}_b$ are correct.
We first show that if $\mathit{Tab}_w[R_w][R_{\overline{w}}] = s$ and hence not $\pm\infty$,
then there exists a set $S_w$ satisfying all the conditions.
For a value of $\mathit{Tab}_w$ to be set to $s$, it means that an update happened in the algorithm,
hence there exist $R_a$ and $R_b$ such that:
$R_{\overline{a}} = can^d_{\overline{V_a}}(R_b \cup R_{\overline{w}})$ ,$R_{\overline{b}} = can^d_{\overline{V_b}}(R_a \cup R_{\overline{w}})$
and $\mathit{Tab}_a[R_a][R_{\overline{a}}] + \mathit{Tab}_b[R_b][R_{\overline{b}}] = s.$
Then we know that there exist $S_a$ and $S_b$ such that $(S_a,R_{\overline{a}})$ $\sigma,\rho$-dominates $V_a$ and
$(S_b,R_{\overline{b}})$ $\sigma,\rho$-dominates $V_b$ and that $|S_a \cup S_b| = s$.
Let $S_w = S_a \cup S_b$, then $S_w$ fulfills the two conditions $|S| = s$ and $R_w \equiv_{V_w}^d S_w$,
now we need to show that $(S_w,R_{\overline{w}})$ $\sigma,\rho$-dominates $V_w$.
Since $(S_b \cup R_{\overline{w}}) \equiv_{V_{\overline{a}}}^d R_{\overline{a}}$ and $(S_a,R_{\overline{a}})$ $\sigma,\rho$-dominates $V_a$ 
it follows from Definition \ref{def_dneigborhood} and \ref{srDomination} that $(S_a,S_b \cup R_{\overline{w}})$ $\sigma,\rho$-dominates $V_a$, 
this is left as an exercise for the reader.
Similarly we conclude that $(S_b,S_a \cup R_{\overline{w}})$ $\sigma,\rho$-dominates $V_b$
Let $S = S_w \cup R_{\overline{w}} = S_a \cup S_b \cup R_{\overline{w}}$ then we get
$S$ $\sigma$-dominates $V_a$ and $S$ $\sigma$-dominates $V_b$, hence it follows
from Definition \ref{srDomination} that $S$ $\sigma$-dominates $V_w$.
Similarly we get $S$ $\rho$-dominates $V_a$ and $S$ $\rho$-dominates $V_b$, hence $S$ $\rho$-dominates $V_w$.
Combining the two last facts it follows from definition \ref{srDomination} that
$(S_w,R_{\overline{w}}) \sigma,\rho$-dominates $V_w$.

Next we will $\forall R_w, R_{\overline{w}}$ show that if there exist an optimal set $S_w \equiv_{V_w}^dR_w$
such that $(S_w,R_{\overline w})$ $\sigma,\rho$-dominates $V_w$,
then $\mathit{Tab}_w[can^d_{V_w}(S_w)][R_{\overline{w}}]~=~|S_w|$.
Let $S_a = S_w \cap V_a$ and $S_b = S_w \cap V_b$.
Since the algorithm goes through all triples of representatives, it will at some point go through $(R_a,R_b,R_{\overline w})$,
where $R_a = can_{V_a}^d(S_a)$ and $R_b = can_{V_b}^d(S_b)$.
Since $(S_w,R_{\overline w})$ $\sigma,\rho$-dominates $V_w$,
$(S_a \cup S_b \cup R_{\overline{w}})$ $\sigma$-dominates $V_w$ and $(S_a \cup S_b \cup R_{\overline{w}}) \rho$-dominates $V_w$.
Then $(S_a,S_b \cup R_{\overline{w}})$ $\sigma,\rho$-dominates $V_a$ and $(S_b,S_a \cup R_{\overline{w}})$ $\sigma,\rho$-dominates $V_b$.
Since in the algorithm $R_{\overline{a}} = can_{V_{\overline{a}}}^d(S_b \cup R_{\overline{w}})$, $(S_a,R_{\overline{a}})$ $\sigma,\rho$-dominates $V_a$.
Similarly we get that $(S_b,R_{\overline{b}})$ $\sigma,\rho$-dominates $V_b$.
Since $S_w$ is the optimal $S_a, S_{\overline{a}}$ and $S_b, S_{\overline{b}}$ must be optimal too,
this means that $\mathit{Tab}_a[R_a][R_{\overline{a}}] + \mathit{Tab}_b[R_b][R_{\overline{b}}] = |S_a \cup S_b| = |S_w| $,
hence $\mathit{Tab}_w[R_w][R_{\overline{w}}] = |S_w|$.

By induction all tables will be correct.
\qed
\end{proof}

\begin{theorem} \label{thm_vs}
For every $n$-vertex graph $G$ given along with a decomposition tree $(T,\delta)$, with $nec_d(T, \delta)$ the maximum $nec(\equiv^d_{V_w})$ of this tree,
any $(\sigma,\rho)$-vertex subset problem on $G$ with $d=d(\sigma,\rho)$ can be solved in time
$O(n^4 \cdot nec_d(T, \delta)^3)$.
\end{theorem}
\begin{proof}
Correctness follows directly from what has been said in this section.
For complexity analysis, for every node $w$ of $T_r$, we basically call the first computation of Lemma~\ref{lem_list_cano_rep} once, 
then loop through every triplet $R_{\overline w}$, $R_a$, $R_b$ of equivalence classes, 
there are at most $nec_d(T, \delta)$ such triplets.
For each triplet we call the second computation of Lemma~\ref{lem_list_cano_rep} three times,
and since $|R_{\overline w}|, |R_a|, |R_b|$ and $\log(nec_d(T, \delta))$ all are at most $n$ 
, we can perform the table update in $O(n^3)$ time.
\qed
\end{proof}

\begin{corollary} \label{cor_vs}
For every $n$-vertex graph $G$ given along with a decomposition tree $(T,\delta)$ for $\cutbool$,
 any $(\sigma,\rho)$-vertex subset problem on $G$ with $d=d(\sigma,\rho)$ can be solved in time
 $O(n^4 \cdot 2^{3d\cdot\cutboolw(T,\delta)^2}))$.
\end{corollary}

The polynomial part of this runtime will be improved in the full version of this paper.
\begin{remark}
For most of the vertex subset problems,
including all problems mentioned after Definition~\ref{def_sigma_rho} (except for those having a parameter $k$),
we have that either $d=1$ or $d=2$.
More precisely, $d=1$ for Max Independent Set, Min Dominating Set, Min Total Dominating Set and Max or Min Independent Dominating Set, and
$d=2$ for Max Strong Stable Set, Max or Min Perfect Code and Min Perfect Dominating Set.
\end{remark}


\subsection{Dynamic programming for vertex partitioning problems}
This section addresses $\exists D_q$ problems (see Definition~\ref{def1}).
We use similar techniques as those for vertex subset problems.
Recall that $d(\mathbb{N}) = 0$, and for every finite or co-finite set $\mu \subseteq \mathbb{N}$, $d(\mu) = 1+min ( max_{x \in \mathbb{N}} x : x \in \mu, max_{x \in \mathbb{N}} x: x \notin \mu )$.
Let $d=d(D_q)=\max_{i,j}d(D_q[i,j])$.

\begin{definition}\label{qdNeighbour}
\emph{Let $G$ be a graph and let $A \subseteq V(G)$ be a vertex subset of $G$.
Two $q$-tuples $(X_1,X_2,...,X_q)$
and $(Y_1, Y_2,...,Y_q)$ of subsets of $A$ are equivalent, denoted by
$(X_1,X_2,...,X_q) \equiv^{q,d}_A (Y_1, Y_2,...,Y_q)$, if
$$\forall i \forall v \in \overline{A} : (|N(v) \cap X_i| = |N(v) \cap Y_i|) \vee (|N(v) \cap X_i| > d \wedge |N(v) \cap Y_i| > d ).$$}
\end{definition}
\begin{lemma}\label{lem_dqequiv}
 $(X_1,X_2,...,X_q) \equiv^{q,d}_A (Y_1, Y_2,...,Y_q)$ if and only if
$\forall i X_i \equiv^d_A Y_i$.
A consequence is that the number of equivalence classes of $\equiv^{q,d}_A$ is at most that of $\equiv^d_{A}$ to the power $q$.
\end{lemma}

The lemma follows directly from Definitions \ref{dNeighbour} and \ref{qdNeighbour}.
In the sequel we will define the values of $Tab_w$ directly indexed by the equivalence classes.
For this we need to first define canonical representatives.
For a node $w$ of $T_r$,
and $ \mathcal{X} = (X_1,X_2,...,X_q) : X_i \subseteq V_w$,
we define $can_{V_w}^{q,d}(\mathcal{X}) = (can^d_{V_w}(X_1),can^d_{V_w}(X_2),...,can^d_{V_w}(X_q))$.

\begin{definition}\label{vpDomination}
\emph{Let $G$ be a graph and $A \subseteq V(G)$.
Let $\mathcal{X} = (X_1,X_2,...,X_q) \in A^q$ and $\mathcal{Y} = (Y_1, Y_2,...,Y_q) \in \overline{A}^q$.
We say that $(\mathcal{X},\mathcal{Y})$ $D_q$-dominates $A$ if for all $i,j$ we have that $(X_j \cup Y_j)\ D_q[i,j]$-dominates $X_i$ (c.f.\ Definition~\ref{srDomination}).}
\end{definition}

\begin{definition}
\emph{For every node $w$ of $T_r$, for every
$\mathcal{X} = (X_1,X_2,...,X_q) \in A^q$ and every $\mathcal{Y} = (Y_1, Y_2,...,Y_q) \in \overline{A}^q$,
let $\mathcal{R}_{\mathcal{X}}=can^{q,d}_{V_w}(\mathcal{X})$ and
$\mathcal{R}_{\mathcal{Y}}=can^{q,d}_{V_w}(\mathcal{Y})$.
We define the contents of $\mathit{Tab}_w[\mathcal{R}_\mathcal{X}][\mathcal{R}_\mathcal{Y}]$ as
$$ Tab_w[\mathcal{R}_\mathcal{X}][\mathcal{R}_\mathcal{Y}] \stackrel{\mathrm{def}}{=} \left\{ \begin{array}{ll}
TRUE & { \begin{array}{ll} \mbox{ if $\exists$ partition $\mathcal{S} = (S_1,S_2,...,S_q)$ of $V_w$ such that:}\\
      \mbox{ $\mathcal{S} \equiv^{q,d}_{V_w} \mathcal{R}_\mathcal{X}$ and $(\mathcal{S},\mathcal{R}_\mathcal{Y})$ $D_q$-dominates $V_w$} \\
      \end{array}
    }\\
FALSE & \textrm{ otherwise.}\end{array} \right.$$}
\end{definition}

The solution to the $\exists D_q$-problem is given by checking if some entry in the table at the root has value $TRUE$.
The computation of the list of all canonical representatives w.r.t.\ $\equiv^{q,d}_{V_w}$ is basically $q$ times the one given in the previous section.
The same situation holds for the computation of a canonical representative from the input of a $q$-tuplet.
Firstly, initialize all values in all tables to $FALSE$.

\medskip

\noindent\textbf{Updating the leaves:~} 
for a leaf $l$ of $T_r$, like before, we abusively denote the vertex of $G$ mapped to $l$ by $l$, and denote $A=\{l\}$.
Firstly, there are $q$ possible classes $l$ could belong to in a $q$-partition of $A$ (recall that empty sets are allowed). 
We call their canonical representatives respectively
$\mathcal{R}_{\mathcal{X}_1}$,
$\mathcal{R}_{\mathcal{X}_2}$,
\dots,
$\mathcal{R}_{\mathcal{X}_q}$.
Secondly, for vertices in $B=V(G) \setminus \{l\}$ note that they are either neighbors of $l$ or not.
Hence we have at most $d+1$ choices (namely $0,1,...,d-1$, $\geq d$) for each of the $q$ partition classes.
(A consequence is that $\mathit{Tab}_l$ has at most $q (d+1)^{q}$ entries.)
For every canonical representative $\mathcal{R}_{\mathcal{Y}}=(Y_1,Y_2,\dots,Y_q)$ w.r.t.\ $\equiv^{q,d}_B$,
we have that $(\mathcal{R}_{\mathcal{X}_i},\mathcal{R}_{\mathcal{Y}})$ $D_q$-dominates $\{ l \}$ 
if and only if
$\forall j |N(l) \cap Y_j| \in D_q[i,j]$.
Accordingly, we perform the following update for every $i$ and for every $\mathcal{R}_{\mathcal{Y}}$:

$\mathit{Tab}_l[\mathcal{R}_{\mathcal{X}_i}][\mathcal{R}_{\mathcal{Y}}]$ is set to be  $TRUE$ if and only if $\forall j\ |N(l) \cap Y_j| \in D_q[i,j]$.

\medskip

\noindent\textbf{Updating the internal node:~} in the following, $\bigcup_q$ denotes the componentwise union of two $q$-tuples.
For a node $w$ with children $a$ and $b$, the algorithm performs the following steps.
For every canonical representative $\mathcal{R}_{\overline{w}}$ w.r.t.\ $\equiv_{\overline{V_w}}^{q,d}$,
for every canonical representative $\mathcal{R}_a$ w.r.t.\ $\equiv_{V_a}^{q,d}$,
and for every canonical representative $\mathcal{R}_b$ w.r.t.\ $\equiv_{V_b}^{q,d}$, do:
\begin{itemize}
\item Compute $\mathcal{R}_w = can^{q,d}_{V_w}(\mathcal{R}_a \bigcup_q \mathcal{R}_b)$,
$\mathcal{R}_{\overline{a}} = can^{q,d}_{\overline{V_a}}(\mathcal{R}_b \bigcup_q \mathcal{R}_{\overline{w}})$,
$\mathcal{R}_{\overline{b}} = can^{q,d}_{\overline{V_b}}(\mathcal{R}_a \bigcup_q \mathcal{R}_{\overline{w}})$ 
\item If $\mathit{Tab}_w[\mathcal{R}_w][\mathcal{R}_{\overline{w}}] = FALSE$ then 
$\mathit{Tab}_w[\mathcal{R}_w][\mathcal{R}_{\overline{w}}] = \mathit{Tab}_a[\mathcal{R}_a][\mathcal{R}_{\overline{a}}] \wedge \mathit{Tab}_b[\mathcal{R}_b][\mathcal{R}_{\overline{b}}]$
\end{itemize}

\begin{theorem}\label{theo_vertex_part}
For every $n$-vertex, $m$-edge graph $G$ given along with a decomposition tree $(T,\delta)$ and an integer $d$.
Let $nec_d(T, \delta)$ be the maximum $nec(\equiv^d_{V_w})$ of any cut defined by this decomposition,
then any $D_q$-problem on $G$, with $d=\max_{i,j}d(D_q[i,j])$, can be solved in time
$O(n^4 \cdot nec_d(T, \delta)^{3q})$.
\end{theorem}

\begin{proof} The complexity analysis is very similar to the one given in Theorem~\ref{thm_vs} and uses the bound in Lemma \ref{lem_dqequiv}.
The correctness proof follows the same style as the proof of Lemma~\ref{lem_correctness_sigma_rho},
Some steps are not explained here because they were explained in Lemma~\ref{lem_correctness_sigma_rho}.

For the correctness, let $a,b$ be the children of $w$ in $T_r$, assume $\mathit{Tab}_a$ and $\mathit{Tab}_b$ are correct.\\
$(\Rightarrow)$ For this direction of the proof we have that $\mathit{Tab}_w[\mathcal{R}_w][\mathcal{R}_{\overline{w}}]=TRUE$.
Then there must exist some $\mathcal{R}_a, \mathcal{R}_b$ such that
$\mathit{Tab}_a[\mathcal{R}_a][\mathcal{R}_{\overline{a}}]=TRUE$ and
$\mathit{Tab}_b[\mathcal{R}_b][\mathcal{R}_{\overline{b}}]=TRUE$,
where
$\mathcal{R}_{\overline{a}} = can^d_{V_a}(\mathcal{R}_b \bigcup_q \mathcal{R}_{\overline{w}})$ and
$\mathcal{R}_{\overline{b}} = can^d_{V_b}(\mathcal{R}_a \bigcup_q \mathcal{R}_{\overline{w}})$.
Hence there exists $\mathcal{S}_a$ partition of $V_a$ and $\mathcal{S}_b$ partition of $V_b$ such that
$(\mathcal{S}_a,\mathcal{R}_{\overline{a}})$ $D_q$-dominates $V_a$
$(\mathcal{S}_b,\mathcal{R}_{\overline{b}})$ $D_q$-dominates $V_b$.
This means that
$\forall i,j : (S_{a_j} \cup R_{\overline{a}_j})\ D_q[i,j]$-dominates $S_{a_i}$ and
$\forall i,j : (S_{b_j} \cup R_{\overline{b}_j})\ D_q[i,j]$-dominates $S_{b_i}$.
It then follows that:
$\forall i,j : (S_{a_j} \cup S_{b_j} \cup R_{\overline{w}_j})\ D_q[i,j]$-dominates $S_{a_i}$ and
$\forall i,j : (S_{a_j} \cup S_{b_j} \cup R_{\overline{w}_j})\ D_q[i,j]$-dominates $S_{b_i}$.
It then follows that:
$\forall i,j : (S_{w_j} \cup R_{\overline{w}_j})\ D_q[i,j]$-dominates $S_{w_i}$.
Which means $(\mathcal{S},\mathcal{R}_{\overline{w}})$ $D_q$-dominates $V_w$.

$(\Leftarrow)$ For this direction of the proof we have that there exists a partition $\mathcal{S} = (S_1,...S_q)$ of $V_w$
such that: $(\mathcal{S},\mathcal{R}_{\overline{w}})$ $D_q$-dominates $V_w$.
This means that $\forall i,j : (S_{w_j} \cup R_{\overline{w}_j})\ D_q[i,j]$-dominates $S_{w_i}$.
Let $\mathcal{S}_a, \mathcal{S}_b$ be the componentwise intersection of $\mathcal{S}_w$ with $V_a$ and $V_b$ respectively.
We then have:
$\forall i,j : (S_{w_j} \cup R_{\overline{w}_j})\ D_q[i,j]$-dominates $S_{a_i}$ and
$\forall i,j : (S_{w_j} \cup R_{\overline{w}_j})\ D_q[i,j]$-dominates $S_{b_i}$.
Hence
$\forall i,j : (S_{a_j} \cup S_{b_j} \cup R_{\overline{w}_j})\ D_q[i,j]$-dominates $S_{a_i}$ and
$\forall i,j : (S_{a_j} \cup S_{b_j} \cup R_{\overline{w}_j})\ D_q[i,j]$-dominates $S_{a_i}$.
Let $\mathcal{R}{\overline{a}} = can^d_{\overline{V_a}} (\mathcal{S}_b \bigcup_q \mathcal{R}_{\overline{w}})$
and   $\mathcal{R}_{\overline{b}} = can^d_{\overline{V_b}} (\mathcal{S}_a \bigcup_q \mathcal{R}_{\overline{w}})$
then
$\forall i,j : (S_{a_j} \cup R_{\overline{a}_j})\ D_q[i,j]$-dominates $S_{a_i}$ and
$\forall i,j : (S_{b_j} \cup R_{\overline{b}_j})\ D_q[i,j]$-dominates $S_{b_i}$.
Let $\mathcal{R}_a = can_{V_a}^d(\mathcal{S}_a)$ and $\mathcal{R}_b = can_{V_b}^d(\mathcal{S}_b)$ then
$\mathit{Tab}_a[\mathcal{R}_a][\mathcal{R}_{\overline{a}}]=TRUE$ and
$\mathit{Tab}_b[\mathcal{R}_b][\mathcal{R}_{\overline{b}}]=TRUE$.
Since the algorithm goes through all triples, it will at some point go through $(\mathcal{R}_a,\mathcal{R}_b,\mathcal{R}_{\overline w})$.
And it will set $\mathit{Tab}_w[\mathcal{R}_w][\mathcal{R}_{\overline{w}}]$ to true, once it is true it will never change.

By induction all tables will be correct.
\qed
\end{proof}

\begin{corollary}\label{cor_vertex_part}
For every $n$-vertex, $m$-edge graph $G$ given along with a decomposition tree $(T,\delta)$ of $\cutbool$,
any $D_q$-problem on $G$, with $d=\max_{i,j}d(D_q[i,j])$, can be solved in time
$O(n^4 \cdot 2^{3qd\cdot\cutboolw(T,\delta)^2})$.
\end{corollary}

The polynomial part of this runtime will be improved in the full version of this paper.

\begin{remark}
For most of the vertex partitioning problems,
including all problems mentioned after Definition~\ref{def1},
we have that either $d=1$ or $d=2$.
More precisely, $d=1$ for $H$-Homomorphism, and
$d=2$ for $H$-Covering and $H$-Partial Covering.
\end{remark}

\section{Conclusion and Perspectives}\label{sec_conclu}
Consider the Hsu-grid 
$HG_{p,q}$ with $p \geq 2$ and $q=p^{\log p}$.
From Lemma~\ref{main_lem_hsu_grid} we have that $\beta w(HG_{p,q})= \Theta(\log p)$ and  $rw(HG_{p,q})=\Theta(p)$.
Since $|V(HG_{p,q})|=p^{1+\log p}$ we get that the runtime of our algorithms (in Theorems \ref{thm_vs} and \ref{theo_vertex_part}) will be polynomial in the input size when expressed as a function of boolean-width but not polynomial when expressed as a function of rank-width.
This assumes that we are given as input not only the graph
$H_{p,q}$ but also an optimal decomposition, of optimal boolean-width in the first case, and 
of optimal rank-width in the second case.

The runtime of our algorithms has the square of the boolean-width as a factor
in the exponent. 
For problems where $d=1$ we can in fact improve this to a factor linear in the exponent \cite{BTVinprep}, but that requires a special focus on these cases. In this paper we have instead focused on a much more general class of vertex subset and
vertex partitioning problems.
However,  one would get runtimes with
an exponential factor linear in boolean-width for all problems considered in this paper, if one could show that
the number of d-neighborhood equivalence classes is no more than the number of 1-neighborhood equivalence classes raised to  some function of $d$.
This question can be formulated as a purely algebraic one as follows: First generalize the concept of boolean sums (1+1=1) to d-boolean-sums
($i+j= \min(i+j,d)$). For a boolean matrix A let $R_d(A)$ be the set of vectors over $\{0,1,...,d\}$ that arise from all
possible d-boolean sums of rows of A. Is there a function $f$ such that  $|R_d(A)| \leq |R_1(A)|^{f(d)}$?

The graphs of boolean-width at most one are exactly the graphs of rank-width one, {\sl i.e.} the distance-hereditary graphs.
What about the graphs of boolean-width two, do they also have a nice characterization?
Is there a polynomial-time algorithm to decide if a graph has boolean-width at most two?
More generally, is there an alternative characterization of the graphs of boolean-width $k$?
Besides, is the bound $\beta w(G)\leq\frac{1}{4}rw(G)^2+\frac{5}{4}rw(G)+\log rw(G)$ tight to a multiplicative factor?

The foremost open problem concerns the computation of the boolean-width of a graph and an optimal
decomposition.
For the moment, one must use the algorithm of Hlin\v{e}n\'y and Oum \cite{HO08} computing the optimal rank decomposition of the graph
in order to have a $2^{2*OPT}$-approximation of an optimal boolean decomposition (c.f.\ Remark~\ref{rem_crazy_approx}).
Note that the runtime of that approximation algorithm is FPT when parameterized by boolean-width.
The best we can hope for is an FPT algorithm computing a decomposition
of optimal boolean-width, but any polynomial approximation would also be nice.

\bibliography{BTV_09}
\bibliographystyle{plain}


\end{document}